\documentclass[twocolumn,english,twocolumn]{revtex4-1}
\usepackage[T1]{fontenc}
\usepackage[latin9]{inputenc}
\usepackage{color}
\usepackage{amsmath}
\usepackage{graphicx}
\usepackage[normalem]{ulem}

\usepackage{babel}

\begin{document}

\title{Bound pulse trains in arrays of coupled spatially extended dynamical systems}

\author{D.$\,$Puzyrev$^{1}$, A. G.$\,$Vladimirov$^{2,3}$, A.$\,$Pimenov$^{2}$,
S. V.$\,$Gurevich$^{4,5}$, S.$\,$Yanchuk$^{1}$}

\affiliation{$^{1}$Institute of Mathematics, Technische Universit{ä}t Berlin,
Strasse des 17. Juni 136, D-10623 Berlin, Germany~~\\
 $^{2}$Weierstrass Institute for Applied Analysis and Stochastics,
Mohrenstrasse 39, D-10117 Berlin, Germany~~\\
 $^{3}$Lobachevsky State University of Nizhni Novgorod, pr.Gagarina
23, Nizhni Novgorod, 603950, Russia~~\\
 $^{4}$Institute for Theoretical Physics, University of M{ü}nster,
Wilhelm-Klemm-Str. 9, D-48149 M{ü}nster, Germany~~\\
 $^{5}$Center for Nonlinear Science (CeNoS), University of M{ü}nster,
Corrensstr. 2, D-48149 M{ü}nster, Germany}
\begin{abstract}
We study the dynamics of  an array of nearest-neighbor coupled spatially
distributed systems each generating a periodic sequence of short pulses.
We demonstrate that unlike a solitary system generating a train of
equidistant pulses, an array of such systems can produce a sequence
of clusters of closely packed pulses, with the distance between individual
pulses depending on the coupling phase. This regime associated with
the formation of locally coupled pulse trains bounded due to a balance
of attraction and repulsion between them is different from the pulse
bound states reported earlier in different laser, plasma, chemical,
and biological systems. We propose a simplified analytical description
of the observed phenomenon, which is in a good agreement with the
results of direct numerical simulations of a model system describing
an array of coupled mode-locked lasers. 
\end{abstract}


\maketitle
Nonlinear temporal pulses and spatial dissipative localized structures
 appear in various optical, plasma, hydrodynamic, chemical, and biological
systems \cite{Rosanov02,Vahed20140016,Clerc2005,Arecchi19991,AkhmedievDS2008,LiobashevskiPRL1996,Lloyd2015,Rotermund19913083,Mikhailov2006,PurwinsDS2010,Lier2013,Suzuki1995,barland12}.
Being well-separated from each other these structures can interact
locally via exponentially decaying tails and, as a result of
this interaction, they can form bound states, known also as ``dissipative
soliton molecules'' \cite{Grelu2012}, characterized by fixed distances
and phase differences between individual structures. Such bound states
can emerge due to the oscillatory character of the  interaction force
which is related to the presence of oscillating tails. Another
scenario  occurs in the case of monotonic repulsive  interaction when
either the pulse tails decay monotonically, or a strong nonlocal repulsive
interaction between the pulses is present. In this case the pulses
tend to distribute equidistantly in time or space leading to periodic pulse trains \cite{Elphick1990,Kutz1998,Nizette200695,PhysRevA.94.063854} which, in contrast to closely packed bound states, exhibit large distances
between the consequent pulses. 

In this Letter we show that even in the case  when the pulses in an
individual system exhibit  strong repulsion, the formation of bound pulse trains  can be achieved by arranging several systems
in an array with nearest-neighbor coupling. As a result, the pulses
interact not only within one system, but also with those in the neighboring 
ones leading to a different balance of attraction and repulsion. More
specifically, we demonstrate that this array can produce a periodic
train of clusters consisting of two or more closely packed pulses
with the possibility to change the interval between the pulses via
the variation of coupling phase parameter. We show that the observed
pulse train states coexist with the regimes which are amplitude synchronized
and possess fixed phase shifts between the pulses emitted by neighboring
array elements. In contrast to the pulse bound state regimes predicted
and observed experimentally  previously \cite{Akhmediev1997,Grelu2012,akhmediev98,Grelu:02,PhysRevA.66.033806,Seong:02,Zhao:07,Purwins1998,7274649,PhysRevE.63.056607,Ortac:10,Wu20113615,Li2012,Gui:13,Tsatourian2013},
this regime cannot exist in a solitary pulse-generating system. We illustrate
this general result by considering a particular example of an array
of mode-locked lasers coupled via evanescent fields in a ring geometry.
Such lasers are widely used for generation of short optical pulses
with high repetition rates and optical frequency combs suitable for
numerous  applications. Combining many lasers into an array one can
achieve much larger output power and substantially improve the characteristics
of the output beam by synchronizing the frequencies of the individual
lasers \cite{Botez_book,1063-7818-33-4-R01,Hagemeier:79,0038-5670-33-3-R03,PhysRevA.46.4252,PhysRevA.49.1301,PhysRevLett.85.3809,KVM01}.
Furthermore, it was recently demonstrated experimentally and verified
theoretically that, in contrast to broad area lasers suffering from
transverse instabilities leading to poor output beam quality, phase
synchronization of individual elements of a multistripe semiconductor
laser arrays can be used to generate high power beams with low far-field
divergence \cite{Jechow09,Lichtner12}.

The correspondence between spatially extended  and time-delay systems
was established in series of publications \cite{Kashchenko1998,Giacomelli1996,Yanchuk2015a,YanchukGiacomelli2017}.
In particular,  it was shown that  delay differential equations (DDEs)
can be reduced to the well known Ginzburg-Landau amplitude equation
in a vicinity of a bifurcation point. On the other hand, many problems
expressed in terms of partial differential equations can be reformulated
in terms of DDEs \cite{PhysRevA.72.033808}. Therefore, for our analysis
it is convenient to assume that  each individual array element generating
a periodic  pulse train is described by a set of DDEs. Then the dynamics
of an array of $N$ such elements can be described by the set of symmetrically coupled systems of nonlinear
DDEs  
\begin{equation}
\frac{d{\vec{u}_{j}}}{dt}={\vec{F}}\left[{\vec{u}_{j}}(t),{\vec{u}_{j}}(t-\tau)\right]+C({\vec{u}_{j-1}}+{\vec{u}_{j+1}}).\label{eq:Model1}
\end{equation}
Here ${\vec{u}_{j}}$,  $j=1,...,N$ is the  state variable describing
the $j$-th system and $C$ is the coupling matrix. We assume that
in the absence of coupling, $C=0$, system generates periodic  pulses
with the period close to the delay time $\tau$. In our simulations
we use a particular model describing a mode-locked laser \cite{PhysRevA.72.033808}.
There, ${\vec{u}}=(A(t),G(t),Q(t))^{T}$, where $A$ denotes the complex
electric field amplitude, whereas $G$ and $Q$ are saturable gain
and loss, respectively. The components of the right hand side vector
function ${\vec{F}}$ are defined by $F_{1}=-\gamma A+\gamma\sqrt{\kappa}RA(t-\tau)$,
$F_{2}=G_{0}-\gamma_{g}G-e^{-Q}(e^{G}-1)|A(t-\tau)|^{2}$, and $F_{3}=Q_{0}-\gamma_{q}Q-s(1-e^{-Q})|A(t-\tau)|^{2}$,
 with $R:=\exp\left[\left(1-i\alpha_{g}\right)G-\left(1-i\alpha_{q}\right)Q\right]/2-i\vartheta$.
Here, the parameter $\gamma$ represents the spectral filtering bandwidth,
$\kappa$ is the attenuation factor describing linear non-resonant
intensity losses per cavity round trip, $G_{0}$ is the pump parameter,
which is proportional to the injection current in the gain region,
$Q_{0}$ is the unsaturated absorption parameter, $\gamma_{g}$ and
$\gamma_{q}$ are the carrier relaxation rates in the amplifying and
absorbing sections, and $s$ is the ratio of the saturation
intensities in these two sections. Though all the parameter values
can vary among different lasers, we assume that this variation is
sufficiently small and consider equal parameters. In what follows
we limit our analysis to the physically meaningful situation when
the lasers are coupled via evanescent fields and, hence, the coupling
matrix $C$ has only a single nonzero element $C_{11}=\eta e^{i\varphi}$,
where $\eta$ is the coupling strength and $\varphi$ is the coupling
phase. 

In the absence of coupling, $\eta=0$, for the chosen parameter
values each laser operates in a stable fundamental passive mode-locking
regime with a single sharp pulse per cavity round trip time \cite{PhysRevA.72.033808}.
This regime corresponds to modulated waves (relative periodic orbits)
with $A_{j}(t)=U(t-\theta_{j})e^{i\omega t+i\upsilon_{j}}$, $G_{j}=G(t-\theta_{j})$,
and $Q_{j}=Q(t-\theta_{j})$, where $U(t)$, $G(t)$, and $Q(t)$
are periodic in time with the period $T$ close to the delay $\tau$,
and arbitrary phase shifts $\theta_{j}$ and $\upsilon_{j}$.

For small coupling $\eta$, the phase shifts $\theta_{j}$ and $\upsilon_{j}$
start evolving slowly in time due to the interaction between the lasers
and, as a result, a synchronized state can be achieved. In particular,
due to the index shift symmetry of the system, solutions are observed,
that are synchronized in the amplitude $|A_{j}|=|A|$ and with the
constant phase shift between the adjacent lasers $\upsilon_{j+1}-\upsilon_{j}=2\pi l/N$, $l=0,\dots,N-1$ \cite{Golubitsky1988,PhysRevLett.85.3809,KVM01,Yanchuk2008a,DHuys2008}.
The simplest types of the synchronized regimes are complete in-phase
synchronization ($l=0$) and anti-phase synchronization ($l=N/2$)
for even number of lasers $N$. Note, that there is also a potentially
interesting ''non-invasive'' case $l=N/4$, for which the coupling
vanishes $A_{j-1}+A_{j+1}=0$. For odd values of $N$, however, the
anti-phase and non-invasive synchronization regimes do not exist.

\begin{figure}
\includegraphics[bb=0bp 0bp 1371bp 621bp,width=1\columnwidth]{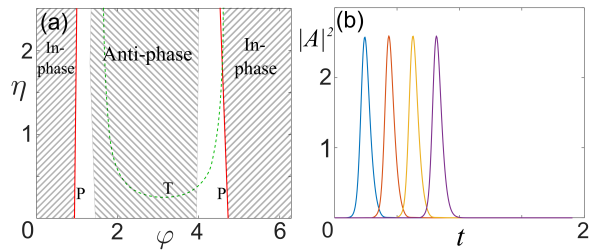}\caption{(a) Bifurcation diagram in coordinates $(\varphi,\eta)$ for the synchronized
solutions in the ring array of four lasers. Left- and right-inclined
hatching indicates the stability regions for the in-phase ($l=0$)
and anti-phase ($l=2$) synchronized solutions. Green line corresponds
to a torus bifurcation (T) and red lines to pitchfork bifurcations
(P) of the in-phase synchronized solution. (b) Laser intensities for
the pulse train bound-state regime in a ring of four lasers calculated
for $\text{\ensuremath{\eta}=0.5}$ and $\varphi=3.0$. Different
colors correspond to different lasers. Other parameters: $\gamma=33.3$,
$\kappa=0.1$, $\alpha_{g}=2.0$, $\alpha_{q}=3.0$, $\vartheta=0$,
$G_{0}=2.0$, $Q_{0}=4.0$, $\gamma_{g}=0.0133$, $\gamma_{q}=1$,
$s=25$, and $\tau=1.875$ (similarly to \cite{PhysRevA.72.033808}).
\label{fig:Bif-diag_all}}
\end{figure}

Further we consider the minimal cases of $N=2$ and $N=4$ lasers,
where $N=4$ is the smallest number that allows in-phase, anti-phase,
and non-invasive synchronized solutions. Figure~\ref{fig:Bif-diag_all}(a)
demonstrates the stability regions for the in-phase and anti-phase
synchronized mode-locked solutions of the system of four lasers using
the master stability function approach \cite{pecora1998master} in
the $(\varphi,\eta)$ plane of coupling parameters. The form of
coupling implies that the stability region of the anti-phase synchronized
solution coincides with that of the in-phase synchronized solution
shifted by $\pi$ with respect to the coupling phase angle $\varphi$.
Furthermore, the P and T lines in Fig. \ref{fig:Bif-diag_all} (a)
show bifurcation thresholds of the in-phase synchronized regime ($l=0$).
In particular, the green line (T) indicates a torus bifurcation threshold
whereas the two red lines correspond to pitchfork bifurcations.

\begin{figure}
\includegraphics[bb=20bp 19bp 1348bp 620bp,width=1\columnwidth]{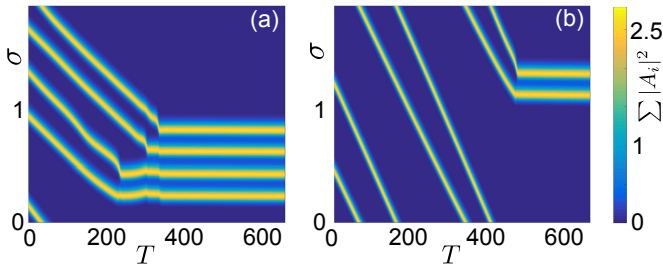}\caption{(a) Space-time diagram of the pulse train boundstate regime for four
lasers in coordinates $(T,\sigma)$, where $\sigma=t\,\mbox{mod\,}\tau_{0}$
is the original fast time and $T=t/\tau_{0}$ the slow time ($\tau_{0}=1.9054$).
Brighter colors indicate higher values of the sum of the laser intensities
$\sum_{i=1}^{4}|A_{i}|^{2}$. (b) Space-time diagram for pulse train
bound state regime for two lasers $(\tau_{0}=1.9043)$. Coupling parameters
are: $\text{\ensuremath{\eta}=0.5}$, $\varphi=3.0$. \label{fig:SeqPulse_pseudo} }
\end{figure}

The torus bifurcation leads to a slight change of the pulse shape
from one pulse period to another, while synchronization and period
of pulsing remains the same. Instead, the pitchfork bifurcations of
the synchronized solution leads to the appearance of a new \emph{bound pulse
train regime}. In this regime, lasers pulse sequentially
on the ring one after another, as shown in Fig.~\ref{fig:Bif-diag_all}(b).
Here, each laser stays close to its fundamental mode-locked regime
with period $\tau_{0}$ close to the delay time $\tau$. The pulse
train bound-state regime can be better visualized using the so-called
pseudo-spatial coordinates plane $(T,\sigma)$ \cite{YanchukGiacomelli2017},
where $\sigma=t\,\mbox{mod\,}\tau_{0}$ is the original fast time
and $T=t/\tau_{0}$ is the slow time (number or round trips, $\tau_{0}=\tau+0.03$),
see Fig.~\ref{fig:SeqPulse_pseudo}(a). We observe that pulses which
were initially distributed on the interval $\sigma\in[0,\tau_{0}]$
start to interact and finally form a bound cluster. The distance between
the pulses in this cluster can be controlled by changing the coupling
phase $\varphi$. 
Similar bound pulse train for the case of two coupled lasers is shown
in Fig.~\ref{fig:SeqPulse_pseudo}(b).
In what follows, we investigate the origin of this bound state solution
by applying the multiscale method \cite{PhysRevE.83.066202,Arkhipov:16}
to the two-laser system in order to find the reduced system of equations
governing the slow dynamics of the time separation between the pulses
and their phase differences.

In order to use the multiscale method, we consider the limit of small
coupling, $\eta=\varepsilon\mu$ with a small parameter $\varepsilon$,
and search for the solution of system (\ref{eq:Model1}) in the form
$A_{j}(t_{0},t_{1})=e^{i\phi_{j}(t_{1})}\mathcal{A}\left[t_{0}+\theta_{j}(t_{1})\right]+\varepsilon A_{j}^{1}(t_{0},t_{1})$,
$G_{j}=\mathcal{G}\left[t_{0}+\theta_{j}(t_{1})\right]+\varepsilon G_{j}^{1}(t_{0},t_{1})$, $Q_{j}=\mathcal{Q}\left[t_{0}+\theta_{j}(t_{1})\right]+\varepsilon Q_{j}^{1}(t_{0},t_{1})$.
Here $\mathcal{A}$, ${\cal G}$, and ${\cal Q}$ is a $\tau_{0}$-periodic
solution of the unperturbed system (mode-locked regime in an uncoupled
laser), $A_{j}^{1}$, $G_{j}^{1}$, $Q_{j}^{1}$ describe first
order corrections due to the coupling between the lasers, $t_{0}=t$
and $t_{1}=\varepsilon t$ are fast and slow times, respectively.

In the following, we explain how the reduced system (\ref{eq:RedSys})
for the the time separation $\Theta=\theta_{2}-\theta_{1}$ between
the pulses and the phase difference $\Phi=\phi_{2}-\phi_{1}$ between
pulses peaks can be obtained. For this purpose, the ansatz above is substituted into (\ref{eq:Model1}) and the resulting system
is expanded in orders of $\varepsilon$ (see \cite{Arkhipov:16,PhysRevE.83.066202}
for more details on this method). In the order $\mathcal{O}(\varepsilon)$,
the following linear system of DDEs for the
vector of perturbations $S_{j}=(\mathrm{\operatorname{Re}}\,A_{j}^{1},\mathrm{\operatorname{Im}}\,A_{j}^{1},G_{j}^{1},Q_{j}^{1})^{T}$
is obtained 
\begin{gather}
-\dot{S}_{j}+a_{1}\left(t\right)S_{j}\left(t\right)+a_{2}(t)S_{j}\left(t-\tau\right)=\label{eq:Eq_NH}\\
a_{3}\dot{\theta}_{j}+a_{4}\dot{\phi}_{j}+\mathcal{R}\left((-1)^{j}\Theta,(-1)^{j}\Phi\right),\nonumber 
\end{gather}
$j=1,2$, with linear operators $a_{1,2}$ and vector functions $a_{3,4}$
depending only on the unperturbed pulse solution. Expressions for
$a_{1,2,3,4}$ and $\mathcal{R}$ are given in the Supplemental material.

The solvability condition (for bounded solutions) of the linear non-homogeneous
system (\ref{eq:Eq_NH}) requires that its right hand side is orthogonal
to the neutral (or Goldstone) modes of the adjoint homogenous system
\cite{guo2013bifurcation}. In the case of small coupling coefficient,
$\eta\ll1$, these modes can be approximated by $\psi_{j}^{\dag}$
and $\xi_{j}^{\dag}$ with $j=1,2$, that are related to the phase
shift and the time-shift invariance of the model equations. These
modes can be found numerically (see, e.g. \cite{Arkhipov:16,PhysRevE.83.066202}).
The orthogonality of the right hand side of (\ref{eq:Eq_NH}) to $\psi_{1,2}^{\dag}$
with respect to the inner product $\int_{0}^{T}\left(a_{3}\dot{\theta}_{j}+a_{4}\dot{\phi}_{j}+\mathcal{R}\left((-1)^{j}\Theta,(-1)^{j}\Phi\right)\right)\psi_{j}^{\dag}(t)dt=0$
leads to the system of two ordinary differential equations 
\begin{eqnarray}
p_{\psi}\dot{\theta}_{1}+q_{\psi}\dot{\phi}_{1} & = & \mu R_{\psi}(\Theta,\Phi),\label{eq:p1}\\
p_{\psi}\dot{\theta}_{2}+q_{\psi}\dot{\phi}_{2} & = & \mu R_{\psi}(-\Theta,-\Phi),\label{eq:p2}
\end{eqnarray}
where coefficients $p_{\psi}$, $q_{\psi}$, and $R_{\psi}$ are given
by the the corresponding scalar products cf. the Supplemental material.
Subtracting equations (\ref{eq:p1}) and (\ref{eq:p2}) from one another,
one obtains the equation for the phase difference $\Phi$ and time
separation of the pulses $\Theta$: 
\begin{equation}
p_{\psi}\dot{\Theta}+q_{\psi}\dot{\Phi}=\mu\left(R_{\psi}(-\Theta,-\Phi)-R_{\psi}(\Theta,\Phi)\right).\label{eq:CoefEq1}
\end{equation}
In the same way, the orthogonality conditions to the modes $\xi_{1,2}^{\dag}$
lead to the equation 
\begin{gather}
p_{\xi}\dot{\Theta}+q_{\xi}\dot{\Phi}=\mu\left(R_{\xi}(-\Theta,-\Phi)-R_{\xi}(\Theta,\Phi)\right).\label{eq:CoefEq2}
\end{gather}
Solving now (\ref{eq:CoefEq1}) and (\ref{eq:CoefEq2}) for $\dot{\Theta}$
and $\dot{\Phi}$, we obtain the reduced system of two ordinary differential
equations for the slow time evolution of $\Theta$ and $\Phi$: 
\begin{gather}
\dot{\Theta}=\eta\cos\left(\Phi+\Delta_{\Theta}\left(\Theta\right)\right)f_{\Theta}\left(\Theta\right),\nonumber \\
\dot{\Phi}=\eta\sin\left(\Phi+\Delta_{\Phi}\left(\Theta\right)\right)f_{\Phi}\left(\Theta\right),\label{eq:RedSys}
\end{gather}
where $f_{\Theta,\Phi}(\Theta)\geq0$. The specific
shape of the right hand side of (\ref{eq:RedSys}) is due to the fact
that the function $R_{\psi}(\Theta,\Phi)$ contains only first Fourier
harmonic in $\Phi$. As a result, the dependence on $\Phi$ is a linear combination of $\sin(\Phi)$ and $\cos(\Phi)$ that can be
represented as (\ref{eq:RedSys}). More details are given in the Supplemental
material.

The bound pulse train states correspond to the fixed points of (\ref{eq:RedSys}). These points lying on the intersection
of nullclines of (\ref{eq:RedSys}) are defined by the condition $\cos\left(\Phi+\Delta_{\Theta}\left(\Theta\right)\right)=\sin\left(\Phi+\Delta_{\Phi}\left(\Theta\right)\right)=0$,
which implies that one of the two conditions should be satisfied,
$\Delta_{\Theta}\left(\Theta\right)=\Delta_{\Phi}\left(\Theta\right)$,
or $\Delta_{\Theta}\left(\Theta\right)=\Delta_{\Phi}\left(\Theta\right)+\pi$.
The first condition corresponds to the saddles of the system (\ref{eq:RedSys}),
while the second equation corresponds either to nodes or to foci.
Figure~\ref{fig:SeqPulse_nullclines}(a) shows intersecting nullclines
of (\ref{eq:RedSys}) in the $(\Theta,\Phi)$ phase plane.
Here, blue filled (unfilled) circles depict stable (unstable) nodes,
red filled (unfilled) circles correspond to stable (unstable) foci,
and blue squares \textendash{} to saddles. All of these equilibria
correspond to pulse bound states in system (\ref{eq:Model1}) with
the same stability properties. Note that a particular case $\Theta=0$
corresponds to the synchronized pulses with the zero time separation,
when the system (\ref{eq:RedSys}) transforms into a single equation
$\dot{\Phi}=\mu C_{\Phi}\sin\Phi$, which admits either in-phase $\Phi=0$
or anti-phase synchronization $\Phi=\text{\ensuremath{\pi}}$ as it
was mentioned above.

Noteworthy, the reduced system (\ref{eq:RedSys}) resembles the equations
governing the slow dynamics of the distance and phase difference between
two interacting dissipative solitons in spatially extended systems
described by generalized complex Ginzburg-Landau equation on an unbounded
domain \cite{PhysRevE.63.056607,PhysRevE.75.045601,PhysRevA.44.6954,PhysRevE.56.6020}.
The case of coupled lasers, however, is distinct in two aspects: (i)
unlike the case of complex Ginzburg-Landau equation the presence of
the phase shifts $\Delta_{\Theta,\Phi}\left(\Theta\right)$ in Eqs.
(\ref{eq:RedSys}) allows for the existence of bound states with the
$\Theta$-dependent phase difference between the pulses different
from $0$, $\pi$, and $\pm\pi/2$, and (ii) instead of a countable
set of equidistant roots, the functions $f_{\Theta,\Phi}\left(\Theta\right)$
have no roots at all, which means that in laser arrays there is a
finite number of bound states which are distributed along the $\Theta$-axis
in a more complex manner.
\begin{figure}
\includegraphics[width=0.95\columnwidth]{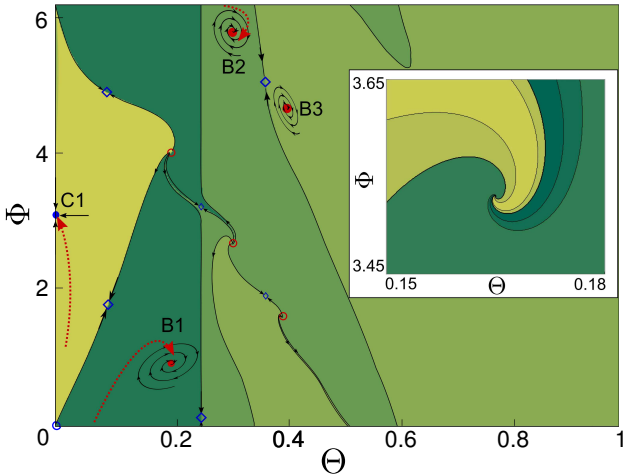}\caption{Stable equilibria and their basins of attraction on the phase plane
of the reduced system (\ref{eq:RedSys}) for coupling phase $\varphi=3.0$.
$\mathbf{C1}$ corresponds to the stable anti-phase synchronized solution.
Equilibria $\mathbf{B1}$, $\mathbf{B2}$, and $\mathbf{B3}$ correspond
to bound states with increasing time separation $\Theta$, which have
different phase shifts $\Phi$ between pulse intensity maxima. Inset:
An example of the intertwining basins of attraction of five stable
bound states in the vicinity of a spiral source for the reduced system
(\ref{eq:RedSys}) for $\varphi=3.99$. \label{fig:SeqPulse_basins} }
\end{figure}
The 2D phase plane of the reduced system (\ref{eq:RedSys})
is presented in Fig.~\ref{fig:SeqPulse_basins}, where the equilibria
and their basins of attraction are shown. Note, that due to the symmetry
$(\Theta,\Phi)\rightarrow(-\Theta,-\Phi)$ it is sufficient to show
only the left half of the coordinate system. Here,
the point $\mathbf{C1}$ corresponds to a stable anti-phase synchronized
solution, while points $\mathbf{B1}$, $\mathbf{B2}$, and $\mathbf{B3}$
indicate the bound states with nonzero pulse time separations $\Theta$. Figure~\ref{fig:SeqPulse_basins}
shows the case of $\varphi=3.0$. For other values of $\varphi$,
there can co-exist from two to five stable equilibria corresponding
to distinct bound states. The basins of attraction of these states
are separated by saddles and, interestingly, they can wind into spiral
sources as it is shown in the inset of Fig.~\ref{fig:SeqPulse_basins}.
The video showing the position of the equilibria and corresponding
basins of attraction for different values of $\varphi$ is available
in the Supplemental material.

A more detailed stability analysis of the bound state corresponding
to the equilibrium $\mathbf{B1}$ is performed numerically using the
path continuation software DDE-BIFTOOL \cite{Engelborghs:2002:NBA:513001.513002}
applied to Eqs.~(\ref{eq:Model1}). The bifurcation
diagram showing the domain of stability of this bound state is presented
in Fig.~\ref{fig:SeqPulse_nullclines}(b). Here, red line P corresponds
to a subcritical pitchfork bifurcation from the in-phase synchronized
solution, whereas the blue F line corresponds to a fold bifurcation
leading the appearance of unstable bound state solutions. The dashed
black line T shows the first torus bifurcation of pulse bound state which leads to a slight change of the pulse shapes from one
pulse period to another, while the period of the pulsing remains the
same. 
\begin{figure}
\includegraphics[bb=15bp 14bp 1295bp 548bp,width=1\columnwidth]{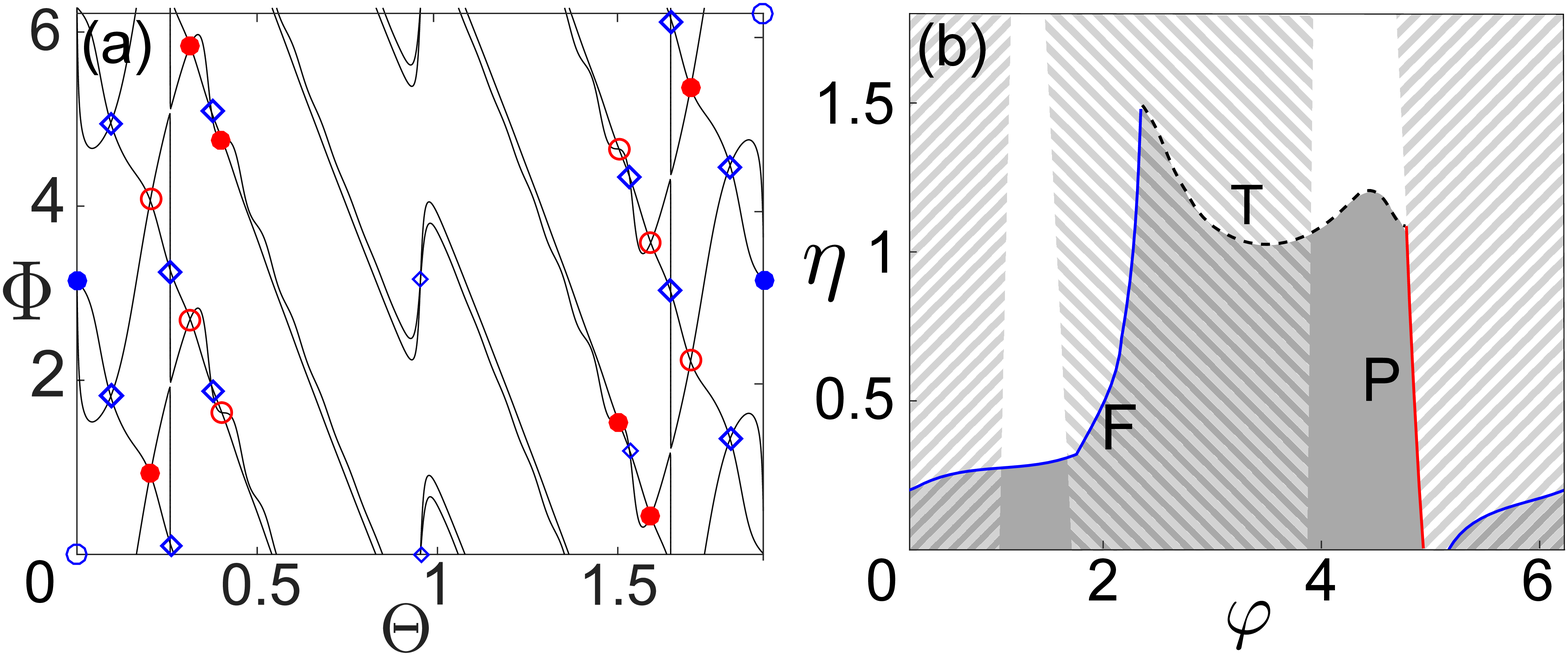}\caption{(a) Fixed points and nullclines for the reduced system (\ref{eq:RedSys})
in the plane $(\Theta,\Phi)$ calculated for coupling phase $\varphi=3.0$.
Blue filled (empty) circles correspond to stable (unstable) nodes,
red filled (empty) circles to stable (unstable) foci, whereas blue
empty diamonds to saddles. (b) Bifurcation diagram for the bound state
$\mathbf{B1}$ (cf. Fig. \ref{fig:SeqPulse_basins}, (b)) in the plane
$(\varphi,\eta)$. Light gray area shows the stability domain of the
bound state. Red line P corresponds to a subcritical pitchfork bifurcation
of the in-phase synchronized solution, blue line F corresponds to
a fold bifurcation, and dashed black line T indicates the first torus
bifurcation. Left- and right-inclined hatching indicates the stability
domains of in-phase and anti-phase synchronized solutions from Fig.
\ref{fig:Bif-diag_all} (a). \label{fig:SeqPulse_nullclines} }
\end{figure}

To conclude, we discovered the bound pulse train regime in an array
of nearest-neighbor coupled nonlinear distributed dynamical systems.
In this regime trains of short pulses generated by individual elements
of the array are bound by local interaction, forming the closely packed
pulse clusters. In the limit of small coupling strength asymptotic
equations are derived governing the slow time evolution positions
and phases of the interacting pulses in an array consisting of two
pulse generators. The pulse separations and phase differences between
the pulses in bound states as well as basins of attraction of different
bound states calculated using this semi-analytical approach are in
good agreement with the results of direct numerical simulations of
a set of DDEs describing an array of coupled
mode-locked lasers (\ref{eq:Model1}). The stability and bifurcations
of bound pulse train regime were studied numerically with the path-following
technique. The bound states reported in this Letter have a similarity
with rather well studied bound states of dissipative solitons in spatially
extended systems, where multiple soliton clusters surrounded by a
linearly stable homogeneous regime can be formed due to a similar
mechanism of balancing between attraction and repulsion. However,
unlike the bound states formed by dissipative solitons, the appearance
of this new type of bound states is related to the presence of coupling
between the neighboring lasers and it is impossible in a solitary
array element, where zero intensity steady state is linearly unstable
and pulse interaction is nonlocal and always repulsive. Furthermore,
unlike the case of complex Ginzburg-Landau-type equations, the new
bound pulse train regime can exhibit continuously changing phase difference
between the pulses depending on their time separation and correspond
to a finite number of fixed points distributed non-equidistantly along
the time axis. Since the physical mechanism of the bound state formation
due to the coupling between neighboring lasers is quite general, it
can be observed in other physical systems described by coupled sets
of partial or delay differential equations, where pulse solutions
are present. Therefore, we believe that our results are generic and
valid for a large class of coupled spatially extended systems of different
physical origin. 
\begin{acknowledgments}
We thank the German Research Foundation (DFG) for financial support
in the framework of the Collaborative Research Center 910, Project
A3 and Collaborative Research Center 787, Project B5. A.V. also acknowledges
the support of the Grant No. 14-41-00044 of the Russian Scientific
Foundation. 

\end{acknowledgments}

\bibliographystyle{apsrev4-1}

\end{document}